\documentclass[a4paper,11pt]{article}
\pdfoutput=1 

\usepackage{jcappub} 

\usepackage[T1]{fontenc} 
\usepackage{subfig}
\usepackage{float}

\usepackage{changes}

\title{\boldmath High-energy gamma-ray emission from SNR G57.2+0.8 hosting SGR J1935+2154}

\author[a,b,1]{Rita C. dos Anjos\note{Corresponding author.},}
\author[c,d]{Jaziel G. Coelho,}
\author[e]{Jonas P. Pereira,} 
\author[f]{Fernando Catalani}

\affiliation[a]{Departamento de Engenharias e Exatas, Universidade Federal do Paran\'a (UFPR), Pioneiro, 2153, 85950-000 Palotina, PR, Brazil}
\affiliation[b]{Applied Physics Graduation Program, Federal University of Latin-American Integration, 85867-670, Foz do Igua\c{c}u , PR, Brazil}
\affiliation[c]{Departamento de F\'isica, Universidade Tecnol\'ogica Federal do Paran\'a, 85884-000 Medianeira, PR, Brazil}
\affiliation[d]{Divis\~ao de Astrof\'isica, Instituto Nacional de Pesquisas Espaciais, Avenida dos Astronautas 1758, 12227-010, S\~ao Jos\'e dos Campos, SP, Brazil}
\affiliation[e]{Nicolaus Copernicus Astronomical Center, Polish Academy of Sciences, Bartycka 18, 00-716, Warsaw, Poland}
\affiliation[f]{Escola de Engenharia de Lorena, Universidade de São Paulo, Estrada Municipal do Campinho, s/n, 12602-810, Lorena, SP, Brazil}

\emailAdd{ritacassia@ufpr.br}
\emailAdd{jazielcoelho@utfpr.edu.br}
\emailAdd{jpereira@camk.edu.pl}
\emailAdd{fcatalani@usp.br}

\abstract{In recent years some efforts have been made to identify active sources capable of accelerating particles up to $10^{15}$ eV, known as PeVatrons. Measurements of TeV ($10^{12}$ eV) gamma-rays from supernova remnants (SNRs) have shown that efficient particle acceleration can occur in SNR diffusive shocks. In this paper, we obtain the contribution to the high energy and very high-energy gamma-ray (VHE, $E > 100$ GeV) emission due to cosmic-ray acceleration from SNR G57.2+0.8 hosting the Soft Gamma Repeater (SGR) J1935+2154 with the use of the GALPROP code. To do so, we take into account the SNR + SGR association as a single source close to the Galactic center. 
We propose that the above setting can provide a more comprehensive scenario for the generation of GeV-TeV gamma-rays. We also discuss the contribution from the SNR G57.2+0.8 and SGR J1935+2154 region to the diffusive TeV energy gamma-ray emission from the Galactic center.}

\begin{document}
\maketitle
\flushbottom

\section{Introduction}

Soft Gamma Repeaters (SGRs) and Anomalous X-Ray pulsars (AXPs) (also known as magnetars) can be possible sources of ultra-high energy cosmic rays (see, e.g.,\cite{2000ApJ...533L.123B,Arons2003}). Rotation and strong magnetic fields can induce large electric potential differences \cite{Aharonian1995, Hooper2009} on their surfaces, accelerating particles up to high energies and filling their atmospheres with electron-positron pairs, which turn into a cascade of particles. Assuming that the magnetospheres of these systems are excellent conductors, the magnitude of electric fields ($E$) is of the order $\omega r B/c$, where $\omega$ is the angular velocity of the star, $B$ its magnetic field, $c$ is the speed of light in the vacuum, and $r$ the distance to the center of the star. Plugging in dipolar magnetic fields, which are reasonable first estimates for the external magnetic field of a star, one obtains $E\sim 2\pi B_0R^3/(Pcr^2)$, where $B_0$ is the magnitude of the magnetic field at the star's surface, and $P$ is its oscillation period. The energy of charged particles such as protons and electrons (with charge magnitude $|e|$) is $\sim 2\pi |e| B_0R^2/(Pc)\sim 10^{18}(B_0/10^{14}\mathrm{G})(R/12\;\mathrm{km})^2(P/1\;\mathrm{s})^{-1}$ eV. Although this is admittedly a rough estimation, it is clear indeed that magnetars could accelerate particles up to ultra-high energies and be of interest for cosmic-ray and gamma-ray telescopes such as HESS, MAGIC and Cherenkov Telescope Array (CTA) \cite{HESS2006, Magic2020, CTA2011}.

Besides inferred surface magnetic fields as high as $10^{14}-10^{15}$ G, SGRs/AXPs are neutron stars (NSs) with the following main properties: spin periods $P\sim(2-12)$ s, spindown rates $\dot{P}\sim(10^{-15}-10^{-10})$  s/s,  persistent X-ray luminosity as large as $10^{35}$ erg/s, transient activity in the form of outbursts of energy around $10^{41}-10^{43}$ erg, and, for some sources, the presence of giant flares of even larger energies, of the order of $10^{44}-10^{47}$ erg (see, e.g., ~\cite{2014ApJS..212....6O} and references therein). They can also present transient events such as glitches, in which $P$ and $\dot P$ suddenly increase. 
The emission nature of SGRs/AXPs is still a reason for debate, and several scenarios have been proposed to explain their observed spectra and properties. Examples thereof are magnetars~\cite{1992ApJ...392L...9D,Thompson1995,2007ApJ...657..967B,2012ApJ...748L..12R,2013ApJ...762...13B}, accreting NSs~\cite{1995A&A...299L..41V,2001ApJ...554.1245A}, rotation-powered pulsars~\cite{2017A&A...599A..87C}, quark stars \cite{2011MNRAS.415.1590O}, or fast-spinning and massive white dwarfs ~\cite{Malheiro2012,Coelho2014c,doi:10.1142/S021827181641025X,2017MNRAS.465.4434C,2020ApJ...895...26B,2020MNRAS.498.4426S}.  For
comprehensive reviews on observations of SGRs/AXPs, see \cite{2008A&ARv..15..225M, 2015RPPh...78k6901T, 2017ARA&A..55..261K}. Given the observational ``shades'' even within the SGR/AXP class, it is beneficial to focus on particular sources for the test of specific physical processes, especially those that are very active or have better measurements.

In this regard, SGR J1935+2154 seems a very interesting candidate. It has been reported for the first time by Swift in 2014 \citep{2014GCN.16520....1S,2014ATel.6294....1C}. Follow-up observations carried out by {\it Chandra} on 2014 July 15 and 29, allowed the precise localization of the source and the detection of its spin period ($P = 3.25$~s), confirming that it is indeed an SGR~\citep{2014ATel.6370....1I}. SGR J1935+2154 has been associated with the middle-aged Galactic supernova remnant (SNR), G57.2+0.8~\citep{2014GCN.16533....1G}, at a possible kinematic distance of $6.6\pm0.7$ kpc~\citep{2011A&A...536A..83S,2013ApJS..204....4P,2020ApJ...905...99Z}. The spin period of SGR J1935+2154 was discovered \cite{2016MNRAS.457.3448I}  through the detection of coherent pulsations (with a period of about $3.24$ s), measured by an extensive observational campaign carried out between 2014 July and 2015 March with {\it Chandra} X-ray Observatory and {\it XMM–Newton}. The source is slowing down at the rate $\dot{P}=1.43(1)\times 10^{-11}$ s/s.
This implies a surface dipolar magnetic field strength of $\sim2.2 \times 10^{14}$ G, a characteristic age $\tau_{\mathrm{age}}\sim3.6$ kyr and a spin-down luminosity $L_{\mathrm{sd}} \sim1.7 \times 10^{34}$ erg/s.

Recently, SGR J1935+2154 has entered a period of unusually intense X-ray burst activity (see ~\cite {2020ApJ...898L..29M,2020ApJ...902L...2B} and references therein). This resulted in the detection of a two-component bright millisecond radio burst on April 28, 2020, similarly to some repeating fast radio bursts (FRBs) observed at extragalactic distances (see \cite{2019A&ARv..27....4P} for a recent review).
To date, two millisecond-duration radio bursts have been detected - the first one by {\it Canadian Hydrogen Intensity Mapping Experiment} (CHIME) and {\it Survey for Transient Astronomical Radio Emission 2} (STARE2) radio telescopes~\citep{2020Natur.587...54C,2020ATel13684....1B}, and the second one by {\textit{Five-hundred-metre Aperture Spherical}} (FAST)~\citep{2020ATel13699....1Z}. These observations support the hypothesis that some FRBs are emitted by magnetars at cosmological distances (see \cite{2020Natur.587...54C,2020NatAs.tmp..232K,2020Natur.587...59B,2021NatAs...5..401T} for details). However, as pointed out in \citep{2020Natur.587...63L}, the connection may be rare.
Many gamma-ray telescopes, including H.E.S.S., also detected an intense SGR J1935+2154 activity on April 28, 2020, and the simultaneous X-ray emission and radio bursts offered the missing link to correlate FRBs with compact objects such as magnetars. The X-ray emission can be an indication that protons and electrons are being accelerated at TeV scales via interactions with matter in the region or through inverse Compton scattering, respectively, and H.E.S.S. observations can help better understand this scenario. If SGRs are cosmic-ray sources at the TeV-PeV energy range, they may contribute to the diffuse gamma-ray emission in the Galaxy given that it is expected to originate from the interactions of cosmic rays with Galactic interstellar gas and radiation fields.

The diffuse gamma-ray emission from the Galactic plane was first detected by EGRET \cite{Hunter_1997} and then followed by Fermi LAT measurements \cite{Fermi2012}. Additionally, a TeV diffuse emission originated in the central part of the Milky Way was detected by ground-based imaging atmospheric Cherenkov telescopes 
\citep {HESS2006, Archer_2016, Magic2020}. 
The observations by H.E.S.S. of a gradient in the cosmic-ray (CR) profile derived from the diffuse VHE gamma-ray emission, with a peak at the inner regions, indicate an injection by a steady source located at the center of the Galaxy \cite{HESS2016}.

In addition, H.E.S.S. has shown the existence of a strong correlation between the distribution of gas clouds in the Central Molecular Zone (CMZ) and the brightness distribution of the diffuse emission, possibly indicating that these gamma-rays are originated from very high-energy protons interacting with matter in these regions. It has been argued that the origin of these cosmic rays via Inverse Compton scattering is probably prevented due to the heavy radiative losses suffered by TeV electrons, restricting their propagation in the CMZ \cite{HESS2018}.

Plus, H.E.S.S. has observed a high-energy cosmic-ray density in the CMZ that is one order of magnitude larger than in the rest of the Galaxy, and that could be explained by the existence of more than one acceleration source in the Galactic Center (GC) \cite{HESS2018}. This clearly motivates studies where G57.2+0.8 and/or SGR J1935+2154 are taken as such sources.
In this year, the Tibet AS$\gamma$ collaboration reported the detection for the first time of sub-PeV diffuse gamma-rays from the Galactic disk. The absence of correlation between the detected events and known TeV sources practically rules out a leptonic origin for this diffuse emission and supports the origin from hadrons being accelerated at PeV energies in that region \cite{Tibet2021}.

In our previous work, we have investigated some properties of the GC and we have obtained the spectral energy and the gamma-ray spectral energy distribution from interacting cosmic rays for this region (see \cite{PhysRevD.101.123015} for details). However, we have considered the GC as the sole source of VHE gamma-rays. Given the possibility that magnetars and supernova remnants could also be natural sources of PeV cosmic rays, the opportunity presents itself to extend our studies and check out their relevance and particularities. This is the main motivation of this work.

Our work is organized as follows. 
The environment of SGR J1935+2154 is presented in Section~\ref{environment}. In Section \ref{model}, we describe our CR propagation model. We detail the numerical implementation and the parameters adopted. In Sect. \ref{results}, we report our results for the model and discuss the TeV gamma-ray emission. The main conclusions are summarized in Section \ref{summary}.

\section{SGR J1935+2154 and its environment}\label{environment}

SNR G57.2+0.8 hosts (the magnetar) SGR J1935+2154, which has previously been detected in active states by X-ray and gamma-ray telescopes~\cite{2020ApJ...904L..21Y,2021MNRAS.tmp..771B}. Following its discovery in 2016, SGR J1935+2154 has probably become the most burst-active SGR, emitting a considerable amount of X-ray bursts over the past few years. 
The association of G57.2+0.8 with SGR J1935+2154, located at its geometric center, has only been proposed recently by \citep{2014GCN.16533....1G}. The distance range to SNR G57.2+0.8 remains highly debated even though various methods have been used (see, e.g., \cite{2016MNRAS.460.2008K,2016MNRAS.457.3448I,2017ApJ...847...85Y,2018ApJ...852...54K}). Most of the distance measurements were targeted at SNR G57.2+0.8 but have used the blackbody emission of SGR J1935+2154. When assuming that SGR 1935+2154 and the SNR are indeed physically associated, Zhong et al. have found that the distance to SGR 1935+2154 is $9.0\pm2.5$~kpc (see \cite{2020ApJ...898L...5Z}). However,
Zhou et al. have performed a molecular environment and a multiwavelength study on G57.2+0.8 (see \cite{2020ApJ...905...99Z} and references therein) and they have shown that the SNR is likely associated with a molecular cloud (MC), which helps constrain the distance by comparing the local standard of rest (LSR) velocity with the Galactic rotation curve. In addition, Zhou et al. have revisited the multiwavelength data in order to constrain the SNR properties such as its age and explosion energy (some shared properties with SGR J1935+2154). It is worth mentioning that Mereghetti et al. have reported an independent distance measurement ($3.1-7.2$ kpc) using the dust scattering X-ray halo around SGR J1935+2154 (see \cite{2020ApJ...898L..29M} for details), and it covers the distance $6.6\pm0.7$ kpc used here.

In this work we use the results of Zhou et al. from the multiwavelength study of G57.2+0.8~\citep{2020ApJ...905...99Z}. We consider that the molecular clouds at $V_{\mathrm{LSR}}=6-14$~km/s and $30$~km/s are spatially overlapping the SNR. The physical interaction between the SNR and MCs is mainly built on a single, weak 1720 MHz maser detected at $V_{\mathrm{LSR}}=30$~km/s, as they are regarded as sign posts of shock–cloud interaction. Moreover, the spatial correlation between the inner shell and mid-IR emission, and the existence of an MC connecting to the inner shell at $V_{\mathrm{LSR}}=30$~km/s, provides an additional morphological support (see \cite{2020ApJ...905...99Z} for details). We stress that further high-resolution molecular observation is needed to provide more kinematic evidence and confirm this association. Here, we consider that SGR J1935+2154 has been associated with the middle-aged SNR G57.2+0.8 at a distance of about 6.6 kpc \cite{2020ApJ...905...99Z}.

The relevance of the detection of MCs in gamma-rays is extensively recognized, especially concerning the issue of the origin of cosmic rays (for a review, see e.g., \cite{2015SSRv..188..187S}).
It is worth stressing that a correlation between the diffuse TeV gamma-ray emission and gas density has also been observed from regions of the Galactic disk characterized by the presence of MCs. In fact, the detection of TeV gamma-rays from some SNRs spatially associated with dense MCs supports the idea that such TeV emission has a hadronic origin and that SNRs might indeed be the sources of Galactic cosmic rays (see \cite{2009ecrs.conf...66G}).
In what follows, we show in details our description of the numerical environment.

\section{Description of the numerical environment}\label{model}

An association with Supernova remnant G57.2+0.8-related SGR J1935+2154 and a MC can contribute to diffuse  gamma-ray fluxes from the GC. Powerful relativistic pulsar winds from SGRs/AXPs could be a promising mechanism for particle acceleration up to $10^{15}$ eV. However, they may be ruled out for higher-energy (> $10^{18}$ eV) accelerations because, due to their compactness and strong magnetic fields, particles lose energy by synchrotron radiation before reaching the classical Hillas boundary \cite{Tjus2020}. These sources can generate cosmic rays due to a magnetic field decay with a time-scale of typically several thousand years and, as a result, the emission is of longer duration \cite{Heyl2010}. Here, we assume that the CR spectra are solely derived from SNR G57.2+0.8 and SGR J1935+2154 and the gamma-ray spectrum can be interpreted as a hadronic and leptonic emission.

The premise that SNRs are plausible cosmic-ray sources inside the Galaxy has the greatest chance of surviving. However, an SNR as a source of PeVatrons is still an open issue. A natural supernova explosion provides a total kinetic energy of $\sim 10^{51}$ erg, generating a cosmic-ray luminosity in the Galaxy of approximately $L_{\mathrm{CR}} \approx 10^{41}$ erg/s and contributing to the cosmic-ray spectrum below the knee. On the other hand, the explosion energy budget of SGRs/AXPs in SNRs varies from $10^{50}$ erg to $2\times10^{51}$ erg. We assume an explosion energy of $10^{50}$ erg due to the association between G57.2+0.8 and SGR J1935+2154, which is reasonable for other SNR sources \cite{2020ApJ...905...99Z, refId0}. Therefore, by assuming that a constant fraction $\eta$ of the kinetic energy is converted into protons and nuclei in the range of GeV-PeV energies, we obtain that $\eta$ needs to be larger than 10\% so that one could consider an SNR + SGR as a source of PeVatron particles \cite{BECKERTJUS20161}:

\begin{equation}
L_{\text{CR}} \approx 10^{41} \hbox{\text{erg/s}} \Biggl(\frac{\eta}{0.1}\Biggl)\Biggl(\frac{\dot{n}}{0.02\; \mathrm{yr^{-1}}}\Biggl)\Biggl(\frac{E_{\mathrm{SN}}}{10^{50}\, \mathrm{erg}}\Biggl),
\end{equation}
where $\dot{n} \sim (1/50 - 1/100)$ yr$^{-1}$ is approximately the rate of supernova explosions. The high energy gamma-ray emission from the decay of neutral pions is produced by inelastic CR proton and nuclei collisions with particles of the neighboring interstellar medium (ISM). The gamma-ray flux is proportional to the total mass of the ISM in the cosmic-ray region \cite{Gnatyk2018}.

We perform 3D GALPROP \cite{Strong_1998, Porter_2017, J_hannesson_2018} simulations using its newest version 56 to obtain the distribution of gamma-rays in the Galaxy \cite{Strong_1998}. The spectra of gamma-rays reaching Earth have been obtained by considering propagation effects and interactions with the background radiation. GALPROP solves the transport equation for a given source distribution and boundary conditions. The simulations include energy losses, fragmentation and decay, convection, diffusive and re-acceleration processes. We adopt the diffusion/re-acceleration model. We have extended the code to take into account the contribution of pointlike CR sources (which would be the effective implementation of an SNR+SGR source). The calculations are made in a cartesian grid assuming the Galactic plane as the X-Y plane with the GC located at its origin. The Galactic volume extends up to 20 kpc in the X and Y directions and has a halo height $(h)$ of 8 kpc with respect to the X-Y plane ($Z=0$). 

We consider a diffusion-reacceleration-convection model. The best-fit diffusion coefficient is a scalar function that is homogeneous and isotropic in the Galaxy and depends on the particle rigidity through a power law with an index $\delta$: $D_{xx} = \beta D_{0}(\rho/\rho_0)^{\delta}$, where $D_{xx}$ is the spatial diffusion coefficient, $\rho_0 = 3$ GV, $D_0$ is a diffusion constant, $\beta$ is the velocity in units of the speed of light. To account for the deflection of cosmic rays by the interstellar magnetic field, we make use of cosmic-ray propagation models with an isotropic diffusion coefficient. In addition, we assume a spatial diffusion inversely proportional to the magnetic field turbulent component $D(\rho,z) = D_{xx}\exp (\left |z \right| /z_t)$, where $z_t$ is a characteristic height scale \cite{Evoli_2017}. However, in future models, an inhomogeneous anisotropic diffusion coefficient will be required to understand recent observations of GeV-TeV gamma-ray diffuse emissions. \cite{2020ApJ...892....6L}. 
The description of the synchrotron intensity is done by means of a double-exponential magnetic field model: $ B(r,z) = B_{0}e^{(R_{\odot}-r)/R_{\mathrm{B}}}e^{-\left |z \right |/z_{\mathrm{B}}}$, where $R_{\odot}= 8.5$ kpc is the solar radius, $B_{0} = 5\;\mu$G, $R_{\mathrm{B}} = 6$ kpc and $z_{\mathrm{B}} = 2$ kpc \cite{Strong_1998, Evoli_2017}. In this analysis we do not consider the influence of the spiral magnetic field.

The source term can be expressed as $Q(\textbf{r},p) = n(\textbf{r})q(p)$, where $n(\textbf{r})$ is the spatial distribution and $q(p)$ the injection energy spectrum of accelerated cosmic rays. All injection spectra are in the form $dq(p)/dp \propto p^{-\alpha}$, where $\alpha$ is the spectral index. We assume just the contribution of a continuous source of energetic particles to the spatial distribution at the position of the SNR G57.2+0.8 environment. The cosmic-ray distribution at Earth is an average of cosmic-ray spectra collected over the time cosmic rays diffuse through the Galaxy, $\sim 10^{7}$ years. The normalization is determined using the total energy of the particles with energies above 1 GeV. We have modified the values of diffusion, environment source localization, and spectral index, which were the most significant parameters. In this scenario, convection and re-acceleration do not play a significant role in the transport of cosmic rays.

VHE gamma-ray emission can be either the byproduct of proton (collisions) or electron (inverse-Compton scattering) interactions with matter and radiation, respectively. In extra-galactic models, the connection between the integral flux of GeV-TeV gamma-rays from a source and the cosmic-ray propagation can result in an upper limit to the total CR luminosity \cite{Supanitsky_2013, Anjos_2014}. Driven by this correlation, we concentrate here on the expected contribution of gamma-rays from SNR G57.2+0.8-related SGR J1935+2154 using upper limit calculations. The method established here provides a relationship between a measured upper limit to the integral flux of GeV-TeV gamma-rays of SNR G57.2+0.8 and its cosmic-ray luminosity.
We assume an isotropic cosmic-ray emission. The secondary gamma-ray flux produced by SNR G57.2+0.8 is proportional to its cosmic-ray flux or luminosity. Therefore, the production rate of the gamma-rays per unit volume is conservative and can be written as a function of the cosmic-ray luminosity:
\begin{equation}
L_{\gamma}(E_{\gamma}) = n_{\mathrm{gas}}\frac{W_{CR}}{4\pi d^{2}\langle E_{0}\rangle}K_{\gamma}P_{\gamma}(E_{\gamma})
\label{ul}
\end{equation}
where $n_{\mathrm{gas}}$ is the gas number density, $W_{\mathrm{CR}}$ is the total energy of the accelerated protons, $\langle E_{0}\rangle$ is the mean energy, $E_{\gamma}$ is the gamma-ray energy, and $K_{\gamma}$ and $P_{\gamma}(E_{\gamma})$ are the number of gamma-rays from cosmic-ray propagation and the energy distribution of the gamma-rays measured on Earth, respectively \cite{Supanitsky_2013, Anjos_2014}. Observations by Fermi and H.E.S.S. were used to constrain the gamma-flux from the propagation of cosmic rays. The confidence upper limit is provided by the power-law index $\Gamma = 2.0$ \cite{FERMI2016}. The upper limit of $5.5\times 10^{-13}$cm$^{-2}$s$^{-1}$ for the integral flux in the 1-10 TeV range was obtained by H.E.S.S. at 99.5\% confidence level (CL) \cite{HESS2018}. The upper limit to the gamma-ray flux of SNR G57.2+0.8 calculated by Fermi at 99.5\% CL is $3.1\times 10^{-10}$ in flux units of $\mathrm{ph}\;\mathrm{cm}^{-2}\mathrm{s}^{-1}$ in the 1–100 GeV energy range. Using these upper limits and equation \ref{ul}, the gamma-ray flux is calculated from G57.2+0.8-related SGR J1935+2154. 

\section{Results and Discussions}\label{results} 

We discuss the GeV-TeV gamma-ray emission from the interstellar environment of SNR G57.2+0.8 and SGR J1935+2154. The gamma-ray emission can be the result of hadronic and leptonic processes coming from interactions of CRs accelerated in the region of the molecular cloud. To account for the contribution to the gamma-ray emission from the propagation of CRs for SNR G57.2+0.8 hosting SGR J1935+2154, we model the CR flux due to SNR G57.2+0.8 as a delta-function centered at the geometric center of SNR G57.2+0.8 and SGR J1935+2154. In our models, the CR injection power is $10^{39}\;\mathrm{erg s}^{-1}$. 

\begin{figure}
  \centering
   \subfloat[$\alpha = 2.4$]{\includegraphics[angle=0,width=0.5\textwidth]{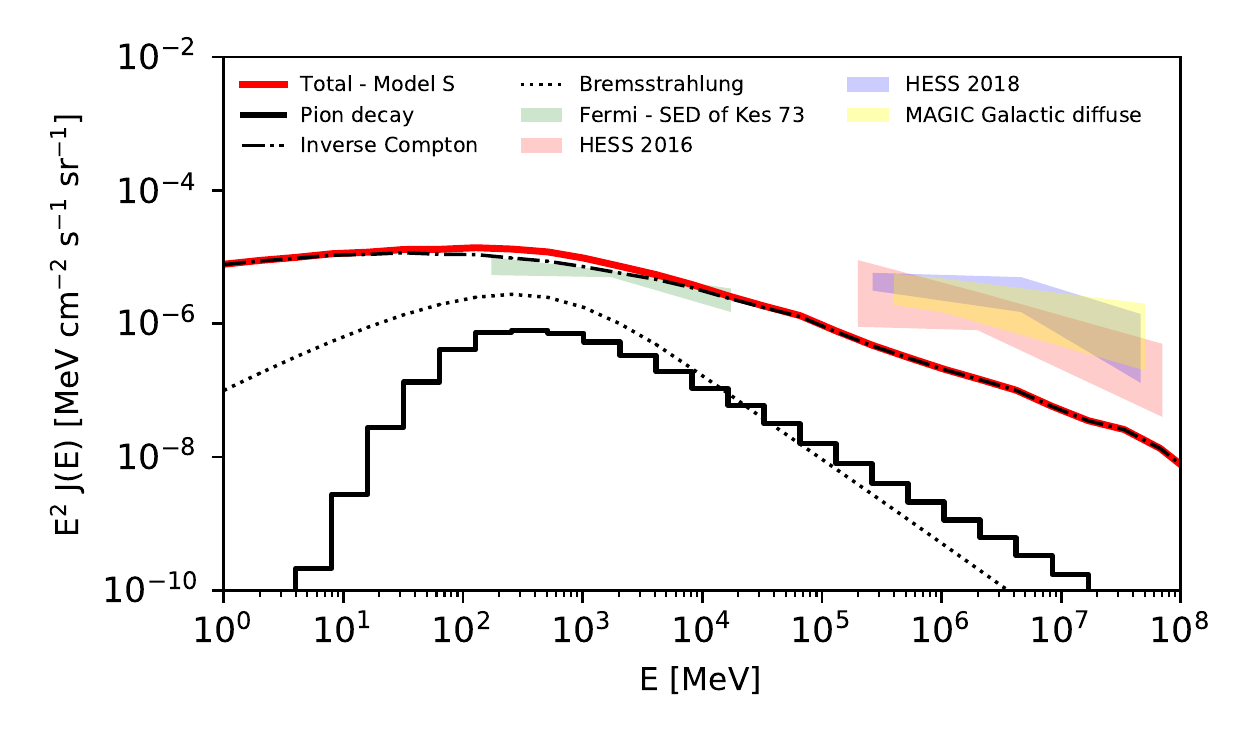}}
    \subfloat[$\alpha = 2.4$]{\includegraphics[angle=0,width=0.5\textwidth]{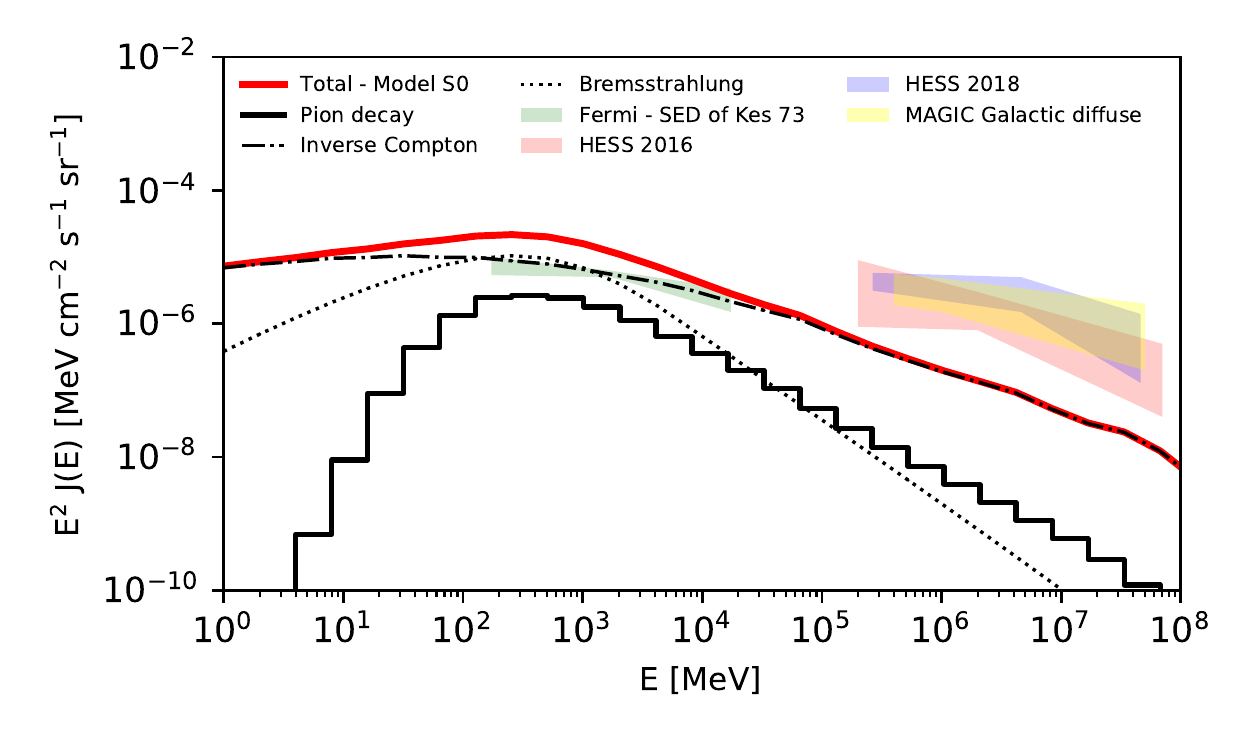}}\\
   \subfloat[$\alpha = 2.0$]{\includegraphics[angle=0,width=0.5\textwidth]{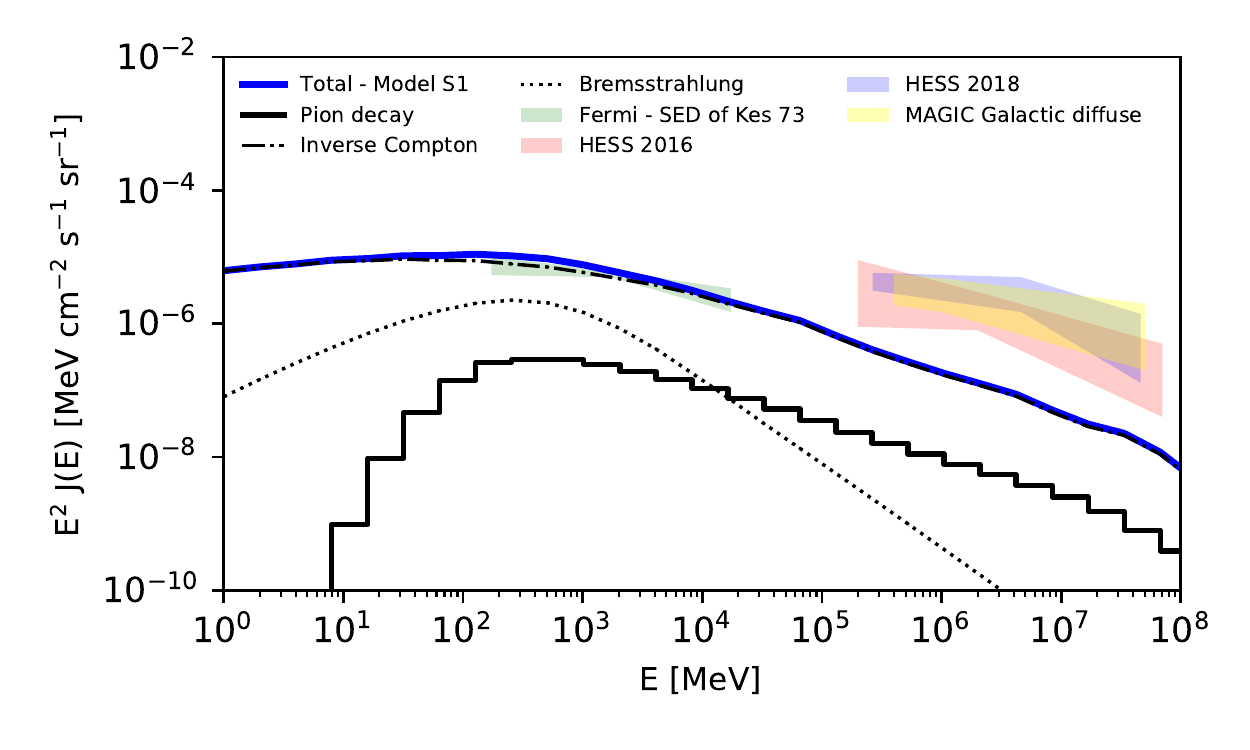}}
    \subfloat[$\alpha = 2.0$]{\includegraphics[angle=0,width=0.5\textwidth]{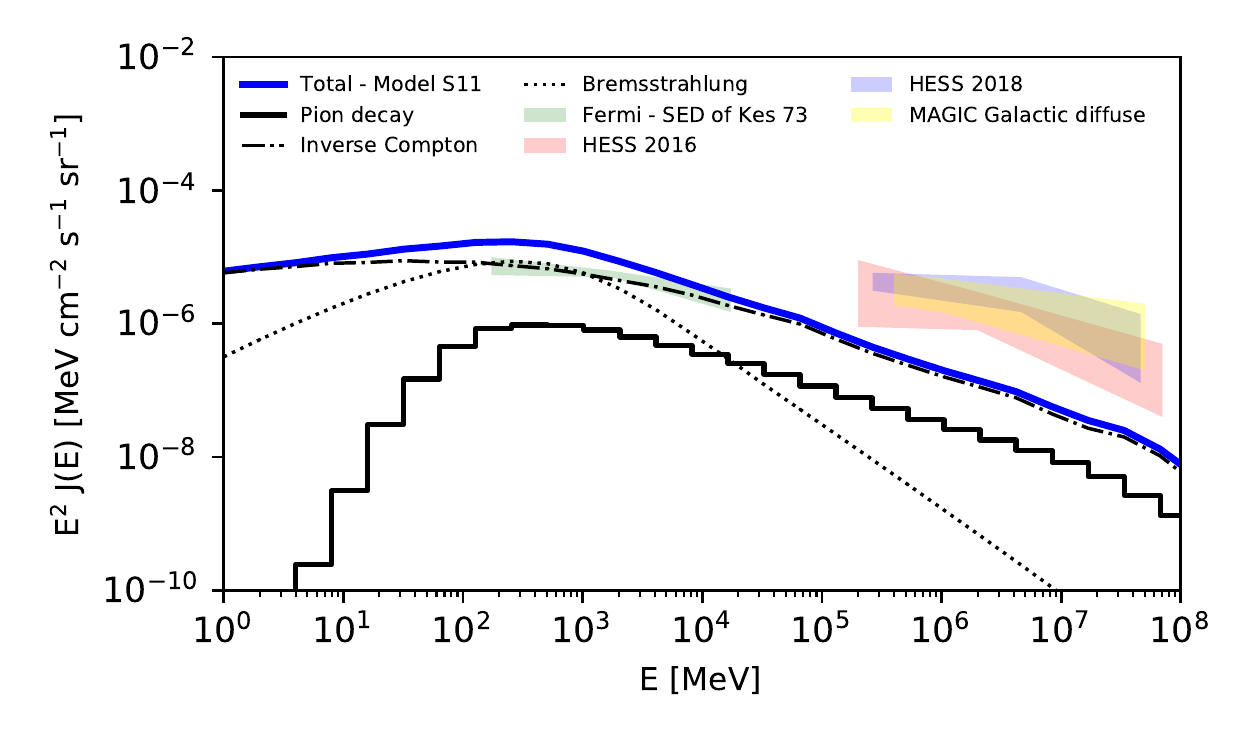}}\\
   \subfloat[$\alpha = 2.5$]{\includegraphics[angle=0,width=0.5\textwidth]{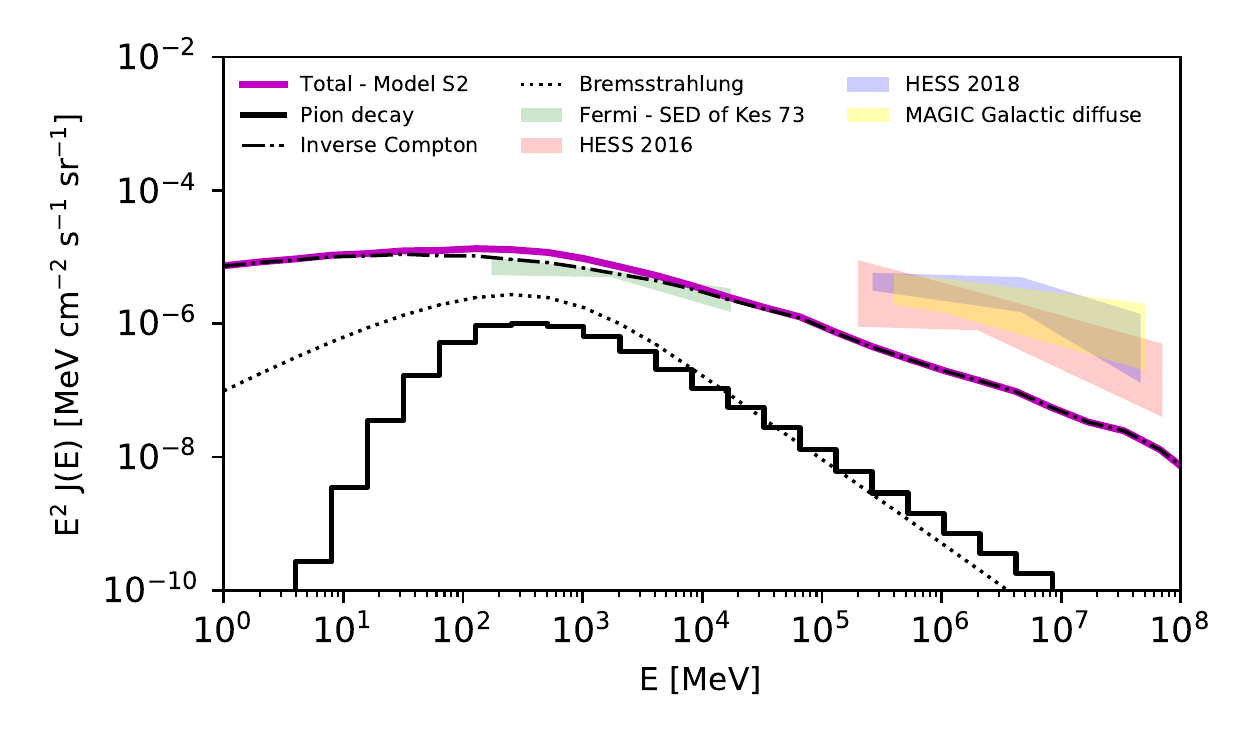}}
   \subfloat[$\alpha = 2.5$]{\includegraphics[angle=0,width=0.5\textwidth]{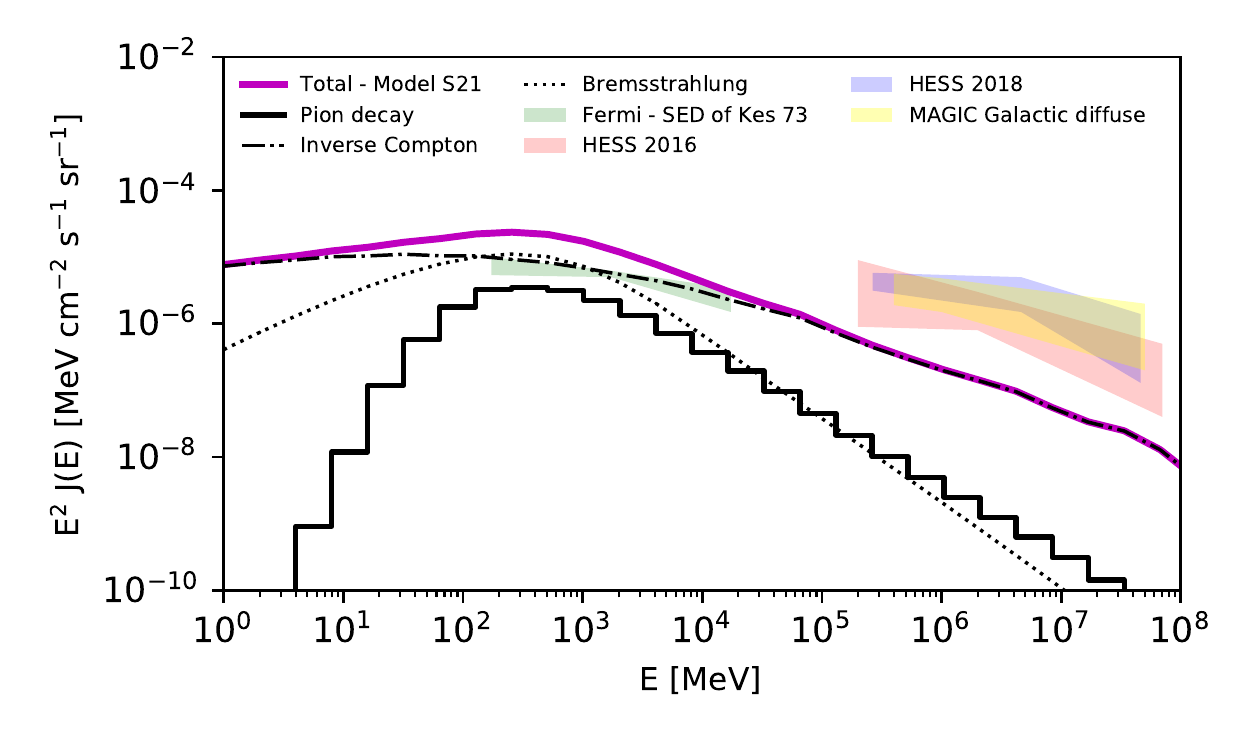}}\\ 
   \caption{Spectral energy distribution of the total gamma-ray emission. The total gamma-ray spectrum is the sum of hadronic and leptonic contributions. Left: Models S, S1 and S2 simulated using distribution of interstellar gas of Fermi-LAT \cite{Fermi2012}. Right: Models S0, S11 and S21 using 2D gas distribution \cite{J_hannesson_2018}.}
    \label{gamma_models}
\end{figure}

Table \ref{tab:1} summarizes the main results of CR source densities for the S models. The choice of parameters was based on our previous work~\cite{PhysRevD.101.123015}. In particular, the value  of the slope of the diffusion coefficient $\delta$ is supported by recent studies, which found that it could have larger values than the expected one for the Kolmogorov turbulence \cite{ 2018A&A...619A..12J, 2008JCAP...10..018E}. Figure \ref{gamma_models} shows the model predictions for the gamma-ray flux. The power-law index varies depending on the species, maximum energy and astrophysical environment considered. We choose a range of values between 2.0 and 2.5 in agreement with the spectrum of individual sources and observations \cite{2001MNRAS.328..393A, 2018A&A...619A..12J}: spectral slope $\alpha$: 2.0 (Models S1 and S11), 2.4 (Models S and S0) and 2.5 (Models S2, S21). The models S, S1 and S2 were simulated using distribution of interstellar gas of Fermi-LAT \cite{Fermi2012} and the models S0, S11 and S21 using 2D gas distribution \cite{J_hannesson_2018}. These gas distributions are currently used to calculate the gamma-ray spectrum from inverse Compton scattering, pion decay and bremsstrahlung, and they have an impact on CR propagation predictions and gamma-ray emission models \cite{Strickland2012, BLUMENTHAL1970BremsstrahlungGases, Porter_2017}.

\begin{table}[h!]
\centering
\caption{GALPROP parameters. See Section~\ref{model} for details.}
\label{tab:1}
\begin{tabular}{|c|c|c|}\hline
Parameter & Unit & Value \\ \hline
   $D_0$ & $\mathrm{cm^{2}s^{-1}}$ & $2.5\times 10^{28}$ \\ 
   $\delta$  & $--$ &  0.6    \\
   Z & $\mathrm{kpc}$ &  8.0  \\    
   X,Y & $\mathrm{kpc}$ &  20.0 \\ 
   $v_{a}$ & $\mathrm{km\;s^{-1}}$   &  28.0 \\  
   $dv/dz$ & $\mathrm{km\;s^{-1}\;kpc^{-1}}$ &  10.0  \\          
   $\alpha$  & $--$ & 2.0 - 2.5  \\ \hline       
 \end{tabular}
\end{table}

\begin{figure}[htb]
  \centering
   \subfloat[Pion Decay - A]{\includegraphics[angle=0,width=0.5\textwidth]{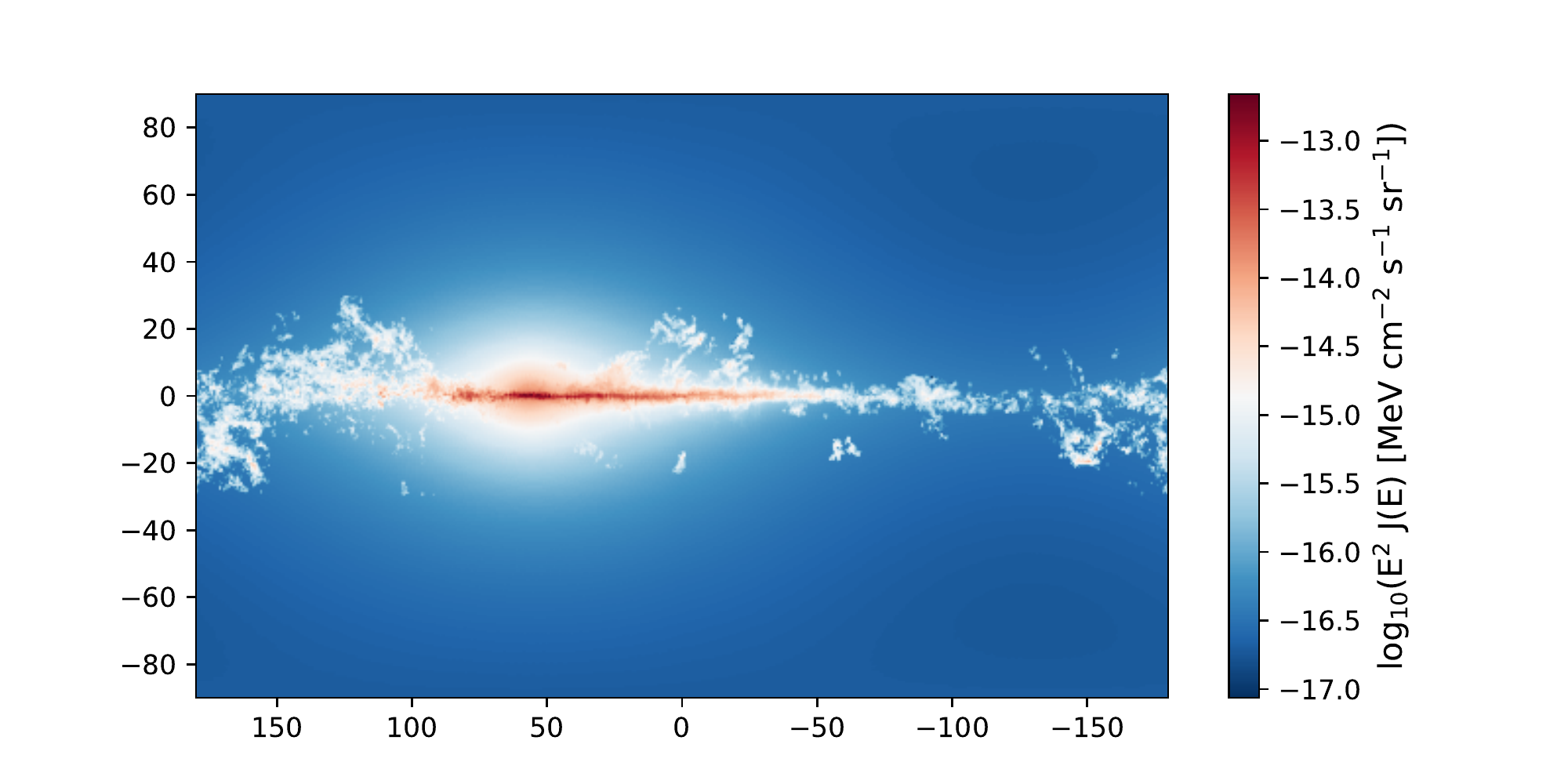}}
   \subfloat[Inverse Compton - A]{\includegraphics[angle=0,width=0.5\textwidth]{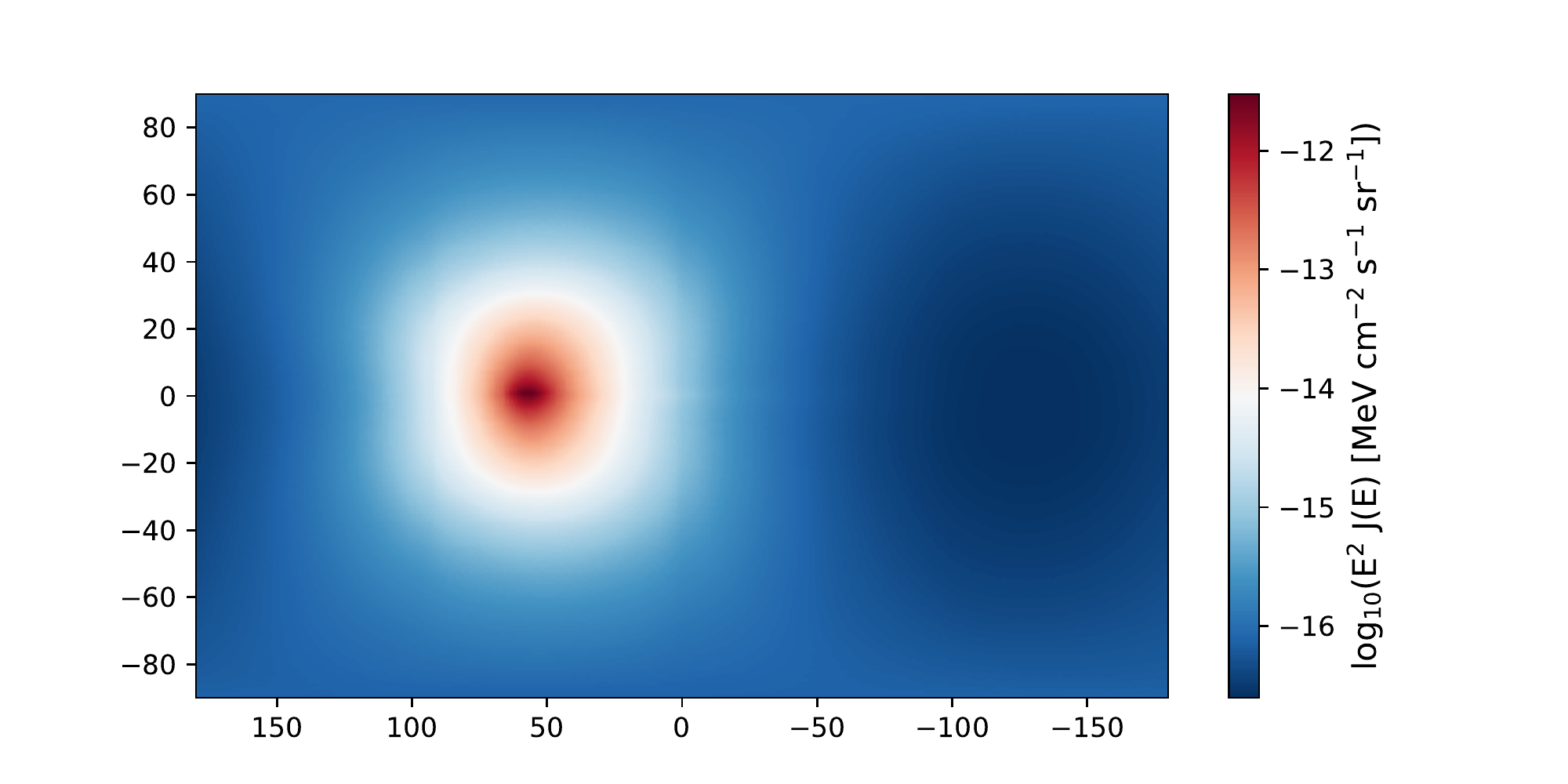}}\\
   \subfloat[Pion Decay]{\includegraphics[angle=0,width=0.5\textwidth]{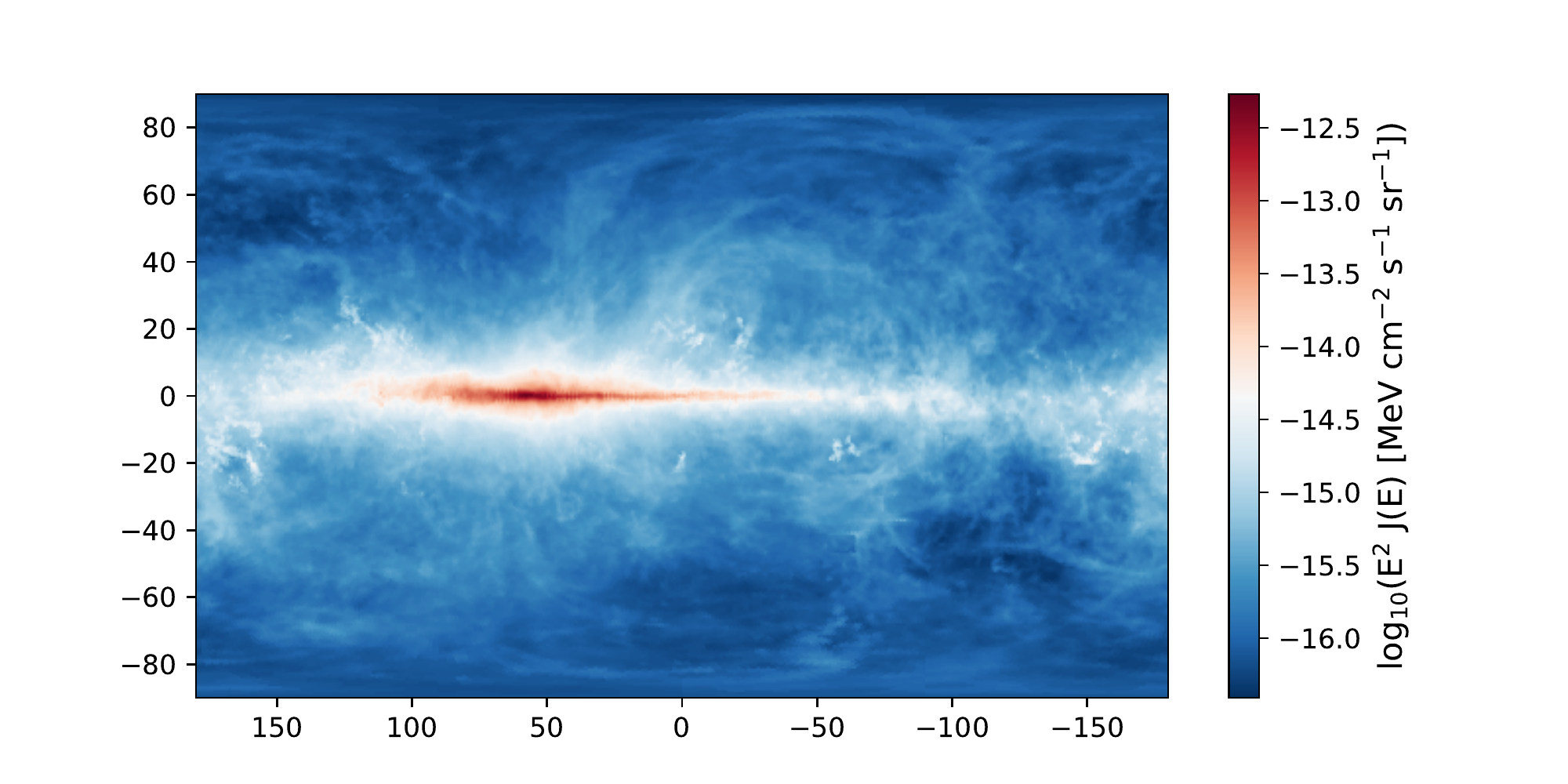}}
   \subfloat[Inverse Compton]{\includegraphics[angle=0,width=0.5\textwidth]{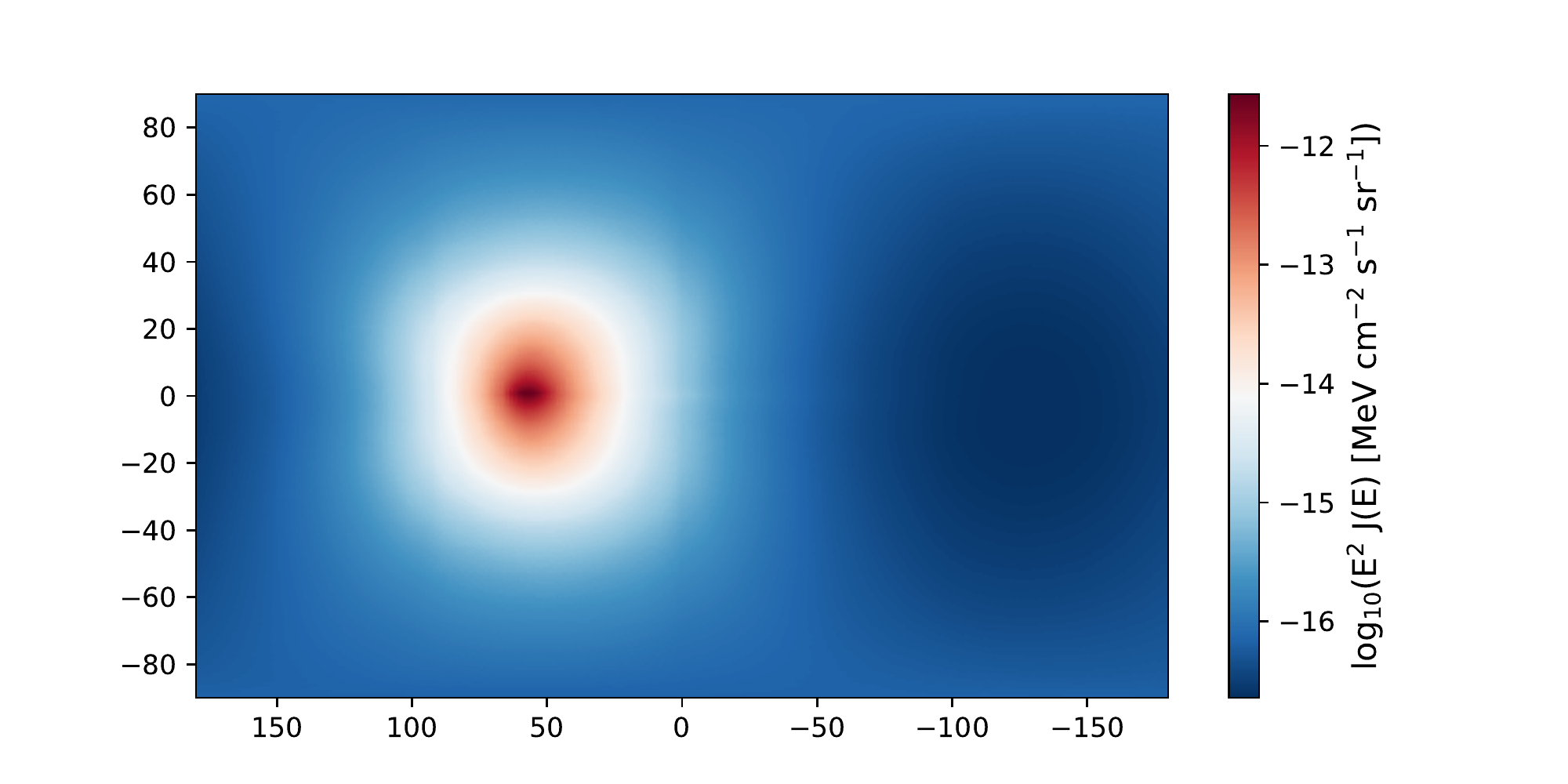}}\\   
    \caption{Diffuse Galactic gamma-rays at 8.0 GeV after nuclei propagation from the SNR G57.2+0.8 and SGR J1935+2154. Hadronic cosmic-ray emission (a,c) and Inverse Compton scattering (b,d) with the distribution of interstellar gas of Fermi-LAT \cite{Fermi2012} - Model A and 2D gas distribution \cite{J_hannesson_2018}. Maps are in Galactic coordinates with $(l,b) = (0,0)$ at the center of the map.}
    \label{diffuse}
\end{figure}

Figure \ref{gamma_models} shows the contribution to the gamma-ray flux received at Earth from the CR flux of SNR G57.2+0.8 and SGR J1935+2154 for the three spectral indices. The assumed gas distribution was the one from the Fermi-LAT collaboration \cite{Fermi2012}. Observations of the diffusive TeV energy gamma-ray emission from the  GC from H.E.S.S. \cite{HESS2016,HESS2018} and MAGIC models \cite{Magic2020} are presented along with a GeV gamma-ray flux emission source in Kes 73 and its molecular environment using Fermi-LAT \cite{Liu_2017} as a reference to determine the reasonableness of the parameters used and to test the model. The young SNR Kes 73 (G27.4+0.0) hosts 1E 1841-045. The model in \cite{Liu_2017} also provides the GeV gamma-ray contribution from this source considering the interaction of the SNR Kes 73 and the surrounding MC. The gamma-ray spectrum obtained can be explained by either a pure hadronic or a hadronic plus a magnetar emission. Therefore, one possible scenario is that SNRs hosting SGRs/AXPs amplify the gamma-ray emission produced by hadronic and leptonic processes and contribute to the diffusive galactic plane emission, in addition to being identified as possible sources of PeVatrons. 

We compute the pion decay, inverse Compton scattering and bremsstrahlung contributions produced by the interactions of the CR with the ISM gas and the radiation field. The inelastic $pp$ collisions produce charged and neutral pions. These are the main mechanisms for the energy loss of protons and nuclei in the ISM. We adopt the interstellar radiation field (ISRF) taken from the most recent official release of GALPROP \cite{J_hannesson_2018}. The hadronic process is the decay of neutral pions into gamma-rays and electron-positrons. In the leptonic process, electrons and positrons lose energy via inverse Compton and synchrotron emissions. In Figure \ref{gamma_models}, one notices that for high energies, the sum of leptonic and hadronic models can collaborate with the observational data from H.E.S.S. and MAGIC model and show the contribution of the SNR and its environment to the diffusive gamma-ray data from the GC, as predicted by analytical forecasts \cite{HESS2016,HESS2018}. The VHE gamma-ray emission from SNR G57.2+0.8 and SGR J1935+2154 was normalized as a sum of pion decay and inverse Compton spectra using the upper limit on the integral flux of TeV gamma-rays from H.E.S.S. at 99.5\% CL \cite{HESS2018}. For higher energies, our analysis suggests that an SNR+SGR may contribute non-negligibly to the total energy density distribution of photons. A precise fraction due to an SNR+SGR is difficult to be obtained due to the extrapolations done and the upper limits used. Notwithstanding, all the above indicates that more precise models should not ignore the contribution from sources other than the GC. The inverse Compton emission component is dominant up to $10^9$ MeV. We have considered the gas distribution adopted by the Fermi-LAT Collaboration \cite{Fermi2012} [Fig. \ref{gamma_models} - (a,c,e)] and by the new Galactic 2D gas distribution \cite{J_hannesson_2018} [Fig. \ref{gamma_models} - (b,d,f)]. The impact of the different gas distribution maps on the models covers the energy range from $10^2$ MeV up to $10^{4}$ MeV and shows a raising contribution from pion emission, see Fig. \ref{diffuse}. 

Despite using a simplified model for the association between SNR G57.2+0.8 and SGR J1935+2154, we obtain a good description of the gamma-ray flux from this region considering the upper limit on the integral flux of TeV gamma-rays from H.E.S.S. at 99.5\% CL \cite{HESS2018}. However, with future measurements of the mass and localization of the molecular cloud, the intensification of the gamma-ray emission from neutral pion decay will help improve the models and fluxes from this region \cite{GRENIER199573, 2020ApJ...905...99Z}. Figures \ref{diffuse} correspond to the pion decay and inverse Compton emissions at 8.0 GeV from the interactions with the distribution of gas in the Galaxy \cite{Fermi2012}. We remark the intensity of the processes around the SNR G57.2+0.8 environment and nearby to the GC, as also shown in Figure \ref{diffuse}. The effect of gas models on CR propagation was shown in \cite{PhysRevD.101.123015}, as well as simulations of gamma-ray radiation.

\section{Summary}\label{summary}

We have investigated particle acceleration models for SNR G57.2+0.8 hosting SGR J1935+2154 and have predicted their GeV-TeV hadronic gamma-ray emissions. The mechanism we demonstrate here is an indication that improves the understanding of the leptonic and hadronic origins of the gamma-ray emission. As a result, we plan to estimate the contribution of the molecular cloud on particle propagation and analyze significant consequences on the predicted gamma signal from this source, provided a region with dense hadronic interactions \cite{GRENIER199573}. Also, such analysis would show whether the gamma-ray emission around this region is due to leptonic or hadronic processes. H.E.S.S. measurements and studies with CTA at VHE - TeV energies of SGR J1935+2154 and other SGRs would be crucial for constraining magnetar particle acceleration scenarios up to the knee. As a result, they may provide independent properties of the magnetospheric regions of magnetars and hence unveil some of the physics taking place in their outermost regions.

\acknowledgments

We are grateful to the anonymous Referee for the suggestions that helped us improve the manuscript. The research of R.C.A. is supported by Conselho Nacional de Desenvolvimento Cient\'{i}fico e
Tecnol\'{o}gico (CNPq) grant numbers 307750/2017-5 and 401634/2018-3, and Serrapilheira Institute grant number Serra-1708-15022. She also thanks for the support of L'Oreal Brazil, with partnership of ABC and UNESCO in Brazil. We acknowledge the National Laboratory for Scientific Computing (LNCC/MCTI, Brazil) for providing HPC resources of the SDumont supercomputer, which have contributed to the research results reported within this paper. URL: http://sdumont.lncc.br. R.C.A. and F.C. acknowledge FAPESP Project No. 2015/15897-1. J.G.C. is grateful for the support of CNPq (421265/2018-3 and 305369/2018-0). J.P.P. acknowledges the financial support from the Polish National Science Centre grant No. 2016/22/E/ST9/00037. R.C.A. and J.G.C. acknowledge the financial support from the NAPI “Fenômenos Extremos do Universo” of Fundação de Apoio à
Ciência, Tecnologia e Inovação do Paraná.

\bibliographystyle{ieeetr}
\bibliography{references}




\end{document}